\documentclass[preprint2]{aastex}

\newcommand{\G}{G79.29+0.46}
\newcommand{\Msun}{$M_{\odot}$}
\newcommand{\Lsun}{$L_{\odot}$}
\newcommand{\mic}{$\mu$m}
\newcommand{\Myr}{$M_{\odot}$\,yr$^{-1}$}
\newcommand{\kms}{km\,s$^{-1}$}

\usepackage{natbib}

\shorttitle{Multiple shells around \G}
\shortauthors{Jim\'enez-Esteban et al.}

\begin{document}

\title{Multiple shells around \G\ revealed from near-IR to millimeter data}

\author{F.~M.~Jim\'enez-Esteban\altaffilmark{1,2,3}, J.~R.~Rizzo\altaffilmark{4,5} and Aina Palau\altaffilmark{6}}
\altaffiltext{1}{Centro de Astrobiolog\'{\i}a (CSIC-INTA), P.O. Box 78, E-28691, Villanueva de la Ca\~nada, Madrid, Spain}
\email{fran.jimenez-esteban@cab.inta-csic.es}
\altaffiltext{2}{Spanish Virtual Observatory}
\altaffiltext{3}{Saint Louis University, Madrid Campus, Division of Science and Engineering, Avda.~del Valle 34, E-28003 Madrid, Spain}
\altaffiltext{4}{Centro de Astrobiolog\'{\i}a (CSIC-INTA), Laboratorio de Astrof\'{\i}sica Molecular, Ctra.~de Ajalvir km.~4, E-28850 Torrej\'on de Ardoz, Spain}
\altaffiltext{5}{Escuela Superior Polit\'ecnica, Universidad Europea de Madrid, Urb.~El Bosque, E-28670 Villaviciosa de Od\'on, Spain}
\altaffiltext{6}{Institut de Ci\`encies de l'Espai (CSIC-IEEC), Campus UAB, Facultat de Ci\`encies, Torre C-5 parell, E-08193 Bellaterra, Barcelona, Spain}

\begin{abstract}

Aiming to perform a study of the warm dust and gas in the luminous
blue variable star \G\ and its associated nebula, we present infrared
Spitzer imaging and spectroscopy, and new CO
$J$\,=\,2\,$\rightarrow$\,1 and 4\,$\rightarrow$\,3 maps obtained with
the IRAM 30m radio telescope and with the Submillimeter Telescope,
respectively. We have analyzed the nebula detecting multiple shells of
dust and gas connected to the star. Using Infrared
Spectrograph--Spitzer spectra, we have compared the properties of the
central object, the nebula, and their surroundings. These spectra show
a rich variety of solid-state features (amorphous silicates,
polycyclic aromatic hydrocarbons, and CO$_2$ ices) and narrow emission
lines, superimposed on a thermal continuum. We have also analyzed the
physical conditions of the nebula, which point to the existence of a
photo-dissociation region.

\end{abstract}


\section{Introduction}

Luminous blue variable (LBV) stars are massive objects which, as they
evolve from the main sequence, undergo a short period of extremely
high mass loss (up to 10$^{-3}$\,\Myr), sometimes accompanied by
so-called giant eruptions, such as the well-known nineteenth-century
outburst of \objectname{$\eta$ Car} \citep{Smith06b}. This mass-loss
strongly influences the further stellar evolution and leads to the
formation of extended circumstellar nebula \cite[see][for a
  review]{Humphreys94}. \cite{Clark05} reported a total of 12 galactic
LBV stars and 23 LBV candidates. From the whole sample, however, only
20 have nebular circumstellar emission. Such nebulae have both gaseous
and dusty components, and two kinds of geometry, spherical and
bipolar, have been recognized.

Since LBV nebulae are strong emitters in the mid- and far-infrared,
the physical and chemical conditions of the material surrounding LBV
stars can be traced studying their spectral characteristics in this
wavelength range. Thus, a few of these nebulae were spectroscopically
studied with the Infrared Space Observatory ({\it ISO}) satellite,
revealing the presence of solid-state features characteristic of
polycyclic aromatic hydrocarbons (PAHs) and silicate dust in both
forms, amorphous and crystalline
\citep[e.g.][]{Lamers96a,Voors99,Voors00b}. Together with these,
numerous forbidden lines on top of a smooth continuum were detected
\citep[e.g.][]{Lamers96b}.

With the advent of {\it Spitzer Space Telescope} \citep{Werner04},
unprecedented high sensitivity and resolution imaging and spectroscopy
have been possible in the infrared, making feasible the study of LBV
nebulae with much great detail. Thus, recently new works have been
published \citep{Morris08, Smith07, Umana09, Gvaramadze09} based on
these new data, which have represented an important step forward in
the knowledge of this brief evolutionary phase of high-mass stars.
 
The molecular gas in LBV nebulae is also a subject of interest. It may
trace radiatively affected and/or shocked regions, which may tell us
about the evolution of the progenitor star. In addition, the chemical
evolution and the formation of circumstellar molecular gas may also be
determined by molecular studies. Gaseous CO was firstly detected
around the \objectname{AG Car} nebula \citep{Nota02}, and after then
in some other objects \citep{Rizzo08a}. Highly excited ammonia was
also detected in the \objectname{Homunculus Nebula} \citep{Smith06a}.

Concerning the molecular counterpart, the most studied source is
\objectname{G79.29+0.46} \citep[][hereafter Paper\,I]{Rizzo08b}. In
this object, the detection of the CO $J$\,=\,3\,$\rightarrow$\,2 line
surrounding the nebula unveiled the presence of moderately dense
(10$^4$\,--\,10$^5$\,cm$^{-3}$) CO gas, probably affected by a C-shock
at a velocity of $\sim$\,15\,\kms.  This CO-ring-like structure has a
diameter of $\sim$\,3\farcm2, and it is strikingly homogeneous in
shape.

The origin of the nebula of \G\ is not clear. From Very Large Array
({\it VLA}) observations of radio continuum and \ion{H}{1},
\cite{Higgs94} showed the thermal nature of the continuum emission and
suggested that the nebula is an ionized shell of $\sim$\,15\,\Msun\ of
swept-up interstellar material. However, after examining the {\it
IRAS} high-resolution images of \G, \cite{Waters96} concluded that it
is a detached shell due to an epoch of high mass loss
($\sim$\,5$\times$10$^{-4}$\,\Myr) during the red supergiant or LBV
phase, followed by a less intense mass-loss period.

Concerning the central object, it has been spectroscopically
studied in the optical and in the near-infrared wavelength range
\citep{Waters96, Voors00b}. Its LBV classification is supported by the
high luminosity ($>$\,10$^{5}$\,\Lsun), the moderate stellar wind
(94\,--\,110\,\kms), and the current high mass-loss rate
(10$^{-6}$\,\Myr) \citep{Waters96}. Unfortunately, its effective
temperature (T$_{eff}$) is not well-known because it is highly
model dependent. The detection of a \ion{He}{1} emission line near
Br$\alpha$ \citep{Waters96} suggests a T$_{eff}$ well above
10,000\,K. On the other hand, the LBV nature of the central star may
limit the T$_{eff}$ to a maximum of 30,000\,K.

In Paper\,I, a careful discussion about the distance to \G\ was
done. We will assume a distance of 1.7\,kpc throughout all the paper.

The fingerprint of stellar evolution onto the
circumstellar/interstellar medium (ISM) can provide a valuable input
to learn about the processes, which govern the star itself and also
affect the evolution of the surrounding gas and dust. The aim of this
paper is to trace the distribution of warm dust, PAHs, and gaseous CO
in relation with \G\ and its associated nebula. To do this, we present
infrared Spitzer imaging and spectroscopy, and new submillimeter CO
$J$\,=\,2\,$\rightarrow$\,1 and 4\,$\rightarrow$\,3 maps obtained with
the IRAM 30m radio telescope and the Submillimeter Telescope ({\it
  SMT}), respectively. In comparison to Paper\,I, this new data set
improves the angular resolution and the analysis of the physical
conditions.

\section{Spitzer data}

The Spitzer data used in this paper have been obtained from the public
Spitzer archive\footnote{\url{http://archive.spitzer.caltech.edu}}.

\subsection{Imaging}

Two cameras are on board of Spitzer: the InfraRed Array Camera
\citep[IRAC;][]{Fazio04} and the Multiband Imaging Photometer
\citep[MIPS;][]{Rieke04}, which provide imaging capability from 3 to
200\,\mic.

\begin{figure*}
\begin{center}
\includegraphics[width=0.85\textwidth,height=!]{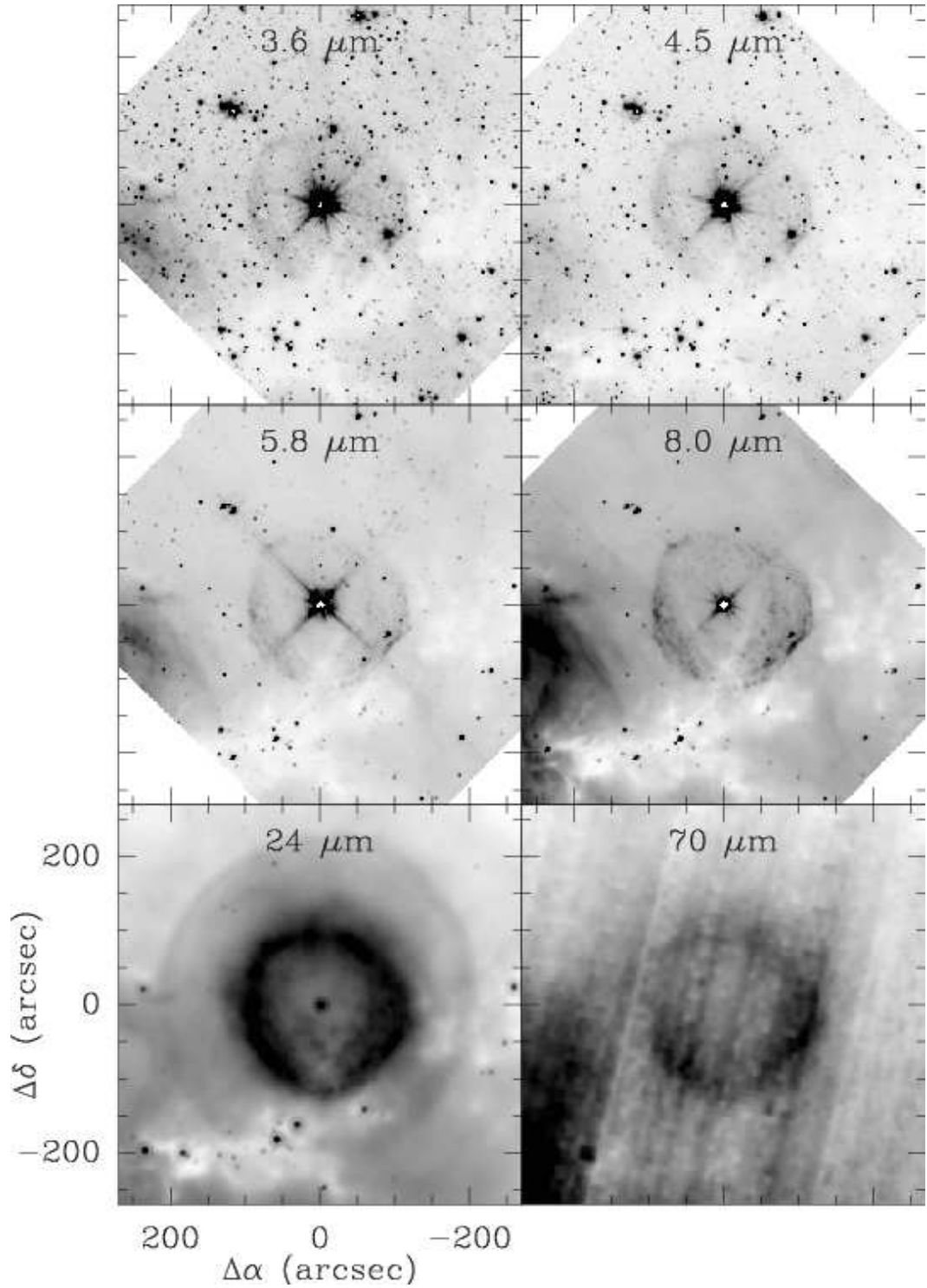}
\caption{Spitzer images of \G. Wavelengths are indicated at the upper
  part of each image, and angular offsets are referred to the central
  star, at (R.A., decl.)$_{J2000} = (20^{h}31^{m}42\fs3, +40\arcdeg
  21\arcmin 58\arcsec)$. A linear brightness scale has been
  artificially selected for each individual image to better show weak
  and diffuse features, instead of point-like objects. Note an outer
  thin second shell clearly visible only at 24\,\mic.
\label{IRAC-MIPS}}
\end{center}
\end{figure*}

Near-infrared images were obtained on 2007 July 05 using the four IRAC
bands at 3.6, 4.5, 5.8, and 8.0\,\mic, as part of the program 30188
(AOR\,17330688). Mid-infrared images at 24 and 70\,\mic\ were also
obtained using MIPS as part of the program 40184 (AOR 22510592) on
2007 December 02. We have worked with the standard post-BCD data sets,
for both IRAC and MIPS images, as retrieved with Leopard from the
Spitzer archive. They were processed with the version S16.1.0 and
S16.1.1,
respectively. Aladin\footnote{\url{http://aladin.u-strasbg.fr}} was
used to cut the images to show a field of view of
9\arcmin$\times$\,9\arcmin\ centered on the LBV star. These images are
shown in Figure~\ref{IRAC-MIPS}. The brightness scale of every filter
has been individually selected to better investigate the nebular
emission of \G.

\subsection{IRS Spectra}

The {\it Spitzer} satellite has a third instrument on board, the Infrared
Spectrograph \citep[IRS;][]{Houck04}. It provides crucial
spectroscopic information in the wavelength range from 5 to 40\,\mic.

\begin{figure*}
\begin{center}
\includegraphics[width=0.85\textwidth,height=!]{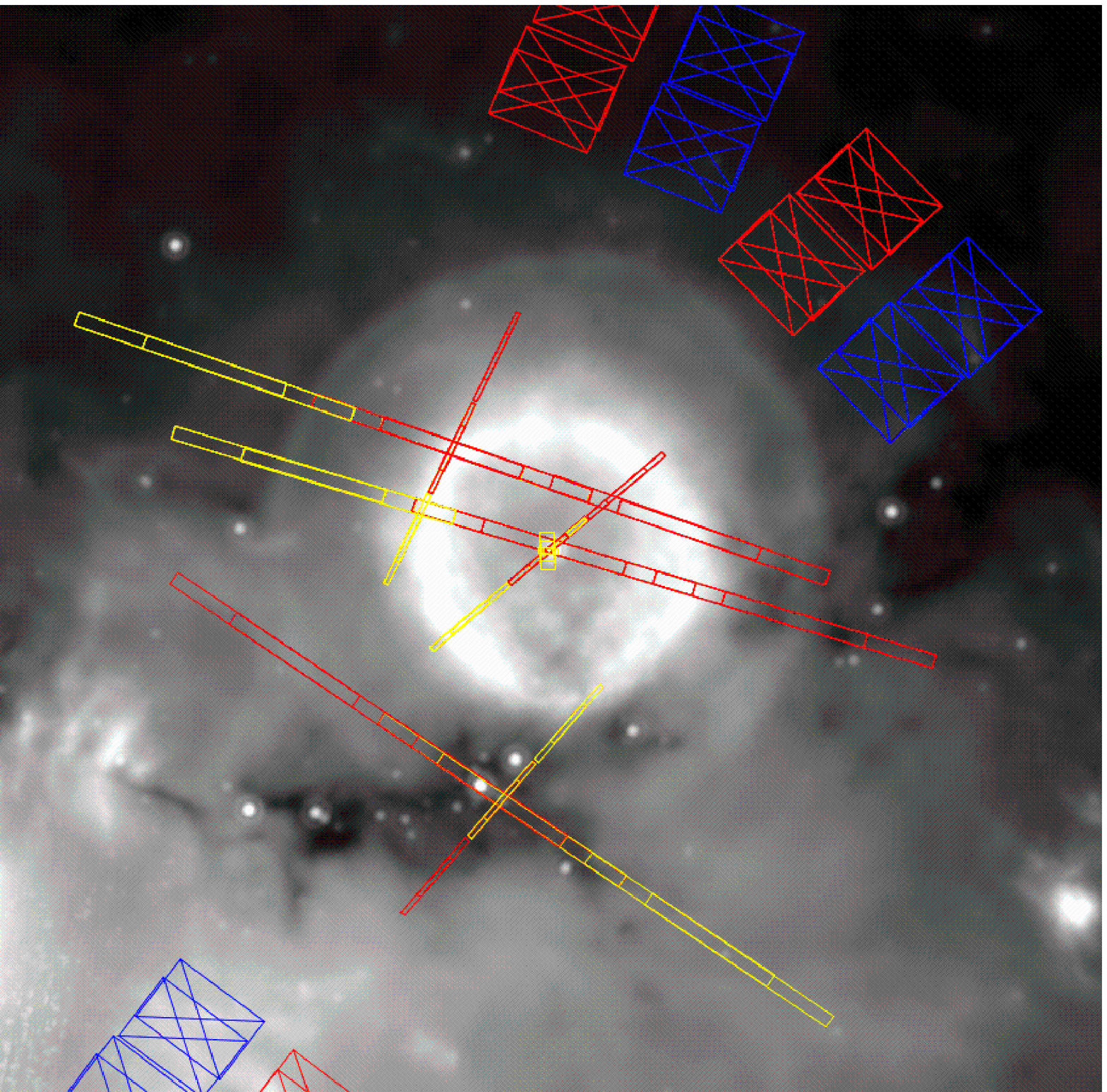}
\caption{Overplot on the MIPS 24\,\mic\ image of the IRS spectral
  slits for all modules, as shown by Leopard software. The field of
  view is 14\arcmin$\times$14\arcmin\ centered on the LBV star. Only
  the observations used in the analysis are shown. The three different
  positions correspond with the central object, the shell, and the
  IRDC. \label{AORs}}
\end{center}
\end{figure*}

Three targets in the field of \G\ have been spectroscopically observed
in four observational programs: a target including the exciting star
and its surroundings (hereafter called `the central object'), a target
centered at a position where the ring nebula is most prominent
(hereafter called `the shell'), and a nearby infrared dark cloud
(hereafter called `the IRDC'). These observations are tabulated in
Table\,\ref{tab:IRSobs}, and the positions of the IRS slits are shown
in Figure~\ref{AORs}. It should be noted that all three, the central
object, the shell, and the IRDC, include emission from nearby regions
than the targets themselves.

\begin{deluxetable}{cccccccccc}
\tabletypesize{\scriptsize}
\tablecaption{IRS observations in the field of \G \label{tab:IRSobs}}
\tablewidth{0pt}
\tablehead{
\colhead{ID} & \colhead{AOR} & \colhead{Date} & \colhead{R.A.} & \colhead{Decl.} & \colhead{Target} & \colhead{Module} & \colhead{$\lambda$ Range} & \colhead{Slit Size} & \colhead{Pipeline} \\
\colhead{program} &  &  & J(2000) & J(2000) &  &  & (\mic) & (arcsec) & 
}
\startdata
  666  &  7557888 & 2003 Nov 14 & $20^{h}~31^{m}~42\fs0$ & $+40\arcdeg\ 21\arcmin\ 42\farcs0$ & The central object  & SL & ~5.2\,--\,14.5 &  3.6$\times$57   & s15.3.0 \\
       &          &            &                       &                                    &           & SH & ~9.9\,--\,19.6 & ~~4.7$\times$11.3 & s15.3.0 \\
       &          &            &                       &                                    &           & LL & 14.0\,--\,38.0 & 10.5$\times$168  & s17.2.0 \\
 1407  &  9761024 & 2004 May 11 & $20^{h}~31^{m}~42\fs1$ & $+40\arcdeg\ 21\arcmin\ 58\farcs8$ & The central object  & SL & ~5.2\,--\,14.5 &  3.6$\times$57   & s15.3.0 \\
       &          &            &                       &                                    &           & LL & 14.0\,--\,38.0 & 10.5$\times$168  & s17.2.0 \\
       &  9761280 & 2004 May 14 & $20^{h}~31^{m}~48\fs9$ & $+40\arcdeg\ 23\arcmin\ 15\farcs4$ & The shell & SL & ~5.2\,--\,14.5 &  3.6$\times$57   & s15.3.0 \\
       &          &            &                       &                                    &           & LL & 14.0\,--\,38.0 & 10.5$\times$168  & s17.2.0 \\
 3121  & 12096768 & 2004 Dec 07 & $20^{h}~31^{m}~45\fs6$ & $+40\arcdeg\ 18\arcmin\ 43\farcs6$ & The IRDC   & SL & ~5.2\,--\,14.5 &  3.6$\times$57   & s15.3.0 \\
       &          &            &                       &                                    &           & LL & 14.0\,--\,38.0 & 10.5$\times$168  & s17.2.0 \\
30188  & 17333248 & 2006 Jun 28 & $20^{h}~31^{m}~42\fs5$ & $+40\arcdeg\ 21\arcmin\ 56\farcs0$ & The central object  & SL & ~5.2\,--\,14.5 &  3.6$\times$57   & s15.3.0 \\
       &          &            &                       &                                    &           & SH & ~9.9\,--\,19.6 & ~~4.7$\times$11.3 & s15.3.0 \\
       &          &            &                       &                                    &           & LH & 18.7\,--\,37.2 & ~11.1$\times$22.3 & s17.2.0 \\
\enddata
\end{deluxetable}

The spectra were obtained by using all possible modules: short--low
(SL), short--high (SH), long--low (LL), and long--high (LH). There are
low-resolution spectra (SL and LL) for the three studied targets,
which cover the spectral range from 5.2 to 38.0\,\mic. However,
high-resolution spectra (SH and LH), covering the spectral range from
9.2 to 37.2\,\mic, are only available for the central object. We have
worked with the standard post-BCD data sets. We have cleaned the
resulting one-dimensional spectra for residual bad pixels flagged by
the pipeline and visually inspected to identify spurious jumps and
glitches. No spurious jumps were detected and the glitches found were
interpolated. Finally, we have merged the different orders of each
module into one final spectrum per module using an IDL code developed
to this end. We have not corrected for basic aperture and slit
loss. Spectra are shown in
Figures~\ref{all-low}\,--\,\ref{Star-high}.

In the case of the central object, the target has been observed on
three different occasions (see Table\,\ref{tab:IRSobs}). The first
time (AOR\,7557888), the IRS slits were not correctly positioned over
the central object. The second time (AOR\,9761024), there should have
been a flux calibration problem, since the two orders do not
match. The third time (AOR\,17333248), the data do not show any
problem. Thus, we have used these last spectra for the analysis,
corresponding with the modules SL, SH, and LH. However, on this last
occasion LL spectrum was not taken, so we have used for the analysis
the LL spectrum corresponding to the AOR\,9761024. The SL spectrum of
the AOR\,9761024 and those of the AOR\,7557888 have been used to
confirm the detection of the solid spectral features and the emission
lines (see Section~\ref{IR-spec}).

\section{CO observations}

\subsection{CO $J$\,=\,2\,$\rightarrow$\,1 Line Emission \label{IRAM}}

We used the IRAM\,30m radio telescope at Pico Veleta (Spain) in 2008
February, to map the CO $J$\,=\,2\,$\rightarrow$\,1 line emission in a
region of $\sim$\,7\arcmin$\times$\,7\arcmin\ around \G. The front end
was the nine-beam HERA receiver \citep{Schuster04}. In order to
optimize both observing time and sampling, we used the {\it
  on-the-fly} mode. A total of four individual maps, in orthogonal
sampling directions, were obtained and later combined in a single data
cube using the GILDAS package. The VESPA autocorrelator was used as
back end, in frequency switching mode. The final bandwidth was 44\,MHz
(equivalent to $\sim$\,57\,\kms\ at the rest frequency), and the
spectral resolution was 80\,kHz ($\sim$\,0.1\,\kms). Due to the
frequency switching, mesospheric CO emission had to be removed from
all the spectra. After folding, low-order polynomials were fitted to
remove baselines using CLASS software package (also included in
GILDAS). Typical rms was 0.2\,K, in $T_{\rm A}^*$ scale.

Calibration was performed during the observations using two absorbers
at different temperatures and later checked up on standard sources
\citep{Mauersberger89}. The uncertainty in calibration is below
15\%. The angular resolution of the radio telescope at 230\,GHz
half-power beam width (HPBW) is 11\arcsec. However, we have further
degraded the final map down to 18\arcsec, in order to reduce the map
noise and to get the same HPBW corresponding to the CO
$J$\,=\,4\,$\rightarrow$\,3 map.

Some other emission lines were observed during the same observing
session. The results from these lines will be released in a future
paper (J.~R.~Rizzo et al.~2010, in preparation), together with a more
complete analysis of the CO $J$\,=\,2\,$\rightarrow$\,1 results.

\subsection{CO $J$\,=\,4\,$\rightarrow$\,3 Line Emission}

The observations of the CO $J$\,=\,4\,$\rightarrow$\,3 line were made
in 2006 June with the SMT, located on Mount Graham, Arizona. The
telescope is a 10.4\,m diameter primary with a nutating
secondary. Design, optics, and structural aspects of the telescope
were extensively described by \cite{Baars99}. The single-channel
receiver SORAL-490 was used under excellent weather conditions
($\tau_{461\,GHz}$\,$\sim$\,0.4\,--\,0.5).  As the back end, we used one
of the acousto-optical spectrometers (AOSCs), which provides an usable
bandwidth of about 250\,MHz (corresponding to 162\,\kms\ of coverage),
and a spectral resolution of 367\,kHz (0.24\,\kms).

The receiver was calibrated with ambient temperature and a liquid
nitrogen-cooled absorbing load. Calibration of the final data has
uncertainties below 20\,\%. System temperature was measured with the
standard sky/ambient temperature load method
\citep{Kutner81}. Calibrations were performed every 15\,--\,20 minutes
to track variations in atmospheric opacity. Typical system temperature
during the observations was in the range of 1000\,--\,1200\,K,
depending on the elevation and the atmospheric opacity.

The pointing was checked within 2 hr on a planet or other bright
continuum source. Corrections to the pointing model were not more than
4\arcsec\ in either azimuth or elevation. Planet observations were
also used to compute the main-beam efficiency at the observed
frequency.

Only a small area toward the southwest of \G\ was observed, where the
most interesting CO features appear (Paper\,I). Position switching was
employed, using the same reference position for all observed
points. Due to time restriction, the CO $J$\,=\,4\,$\rightarrow$\,3 map
was sampled only by one beam (18\arcsec) in both equatorial
coordinates, and hence it is not fully sampled. The reduction
processes were similar to those of IRAM's data, already described in
Section~\ref{IRAM}.

\begin{figure}
\begin{center}
\includegraphics[width=0.5\textwidth,height=!]{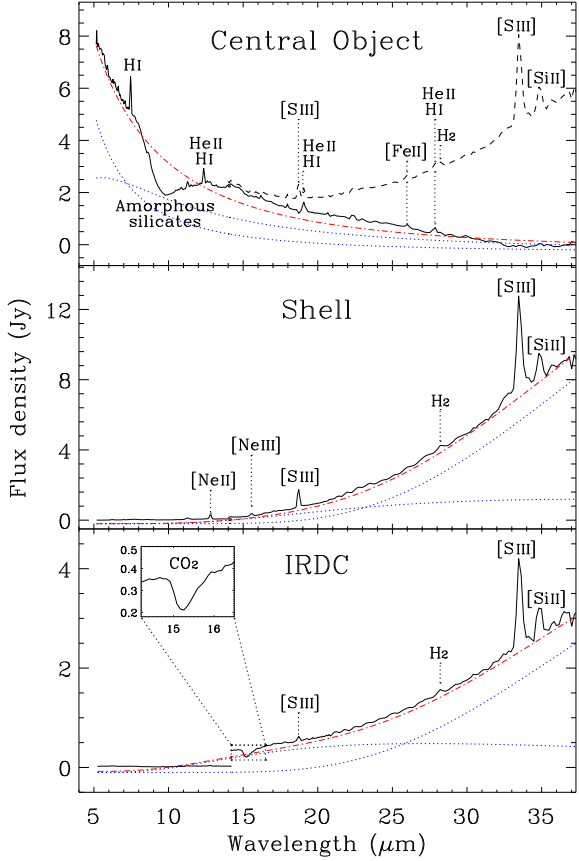}
\caption{IRS low-resolution spectra (modules SL and LL) of the three
  studied targets. The most prominent spectral lines and solid
  features are labeled in the figure. {\em Upper panel}: the spectra
  of the central object after subtracting the LL spectrum of the shell
  (see the text). The dashed line corresponds to the original LL
  spectrum. The dotted (blue) lines correspond to the emission of two
  blackbodies, one at 20,000\,K and the other at 925\,K. The
  dash-dotted (red) line is the combination of these two blackbodies.
  {\em Middle panel}: the spectra of the shell. The dotted (blue)
  lines correspond to the emission of two blackbodies, one at 65\,K
  and the other at 132\,K. The dash-dotted (red) line is the
  combination of these two blackbodies. {\em Lower panel}: the spectra
  of the IRDC. The dotted (blue) lines correspond to the emission of
  two blackbodies, one at 66\,K and the other at 187\,K. The
  dash-dotted (red) line is the combination of these two blackbodies.
\label{all-low}}
\end{center}
\end{figure}

\begin{figure}
\includegraphics[width=0.5\textwidth,height=!]{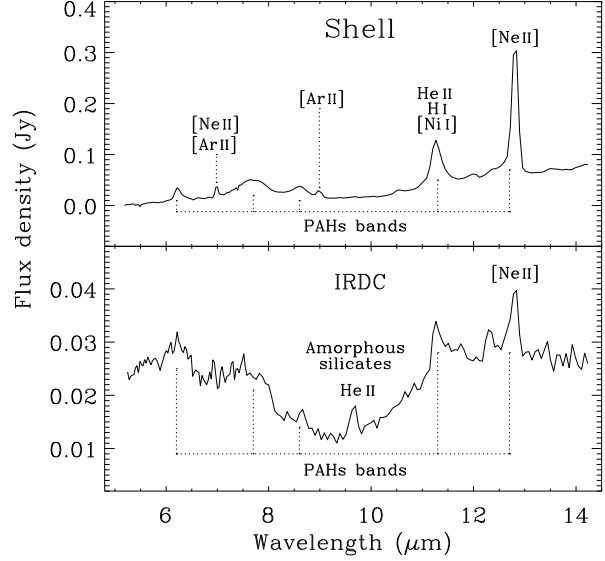}
\caption{IRS SL module spectra of the shell and the IRDC. The
  most prominent spectral lines and solid bands are labeled in the
  figure.
\label{short-low}}
\end{figure}

\begin{figure}
\includegraphics[width=0.5\textwidth,height=!]{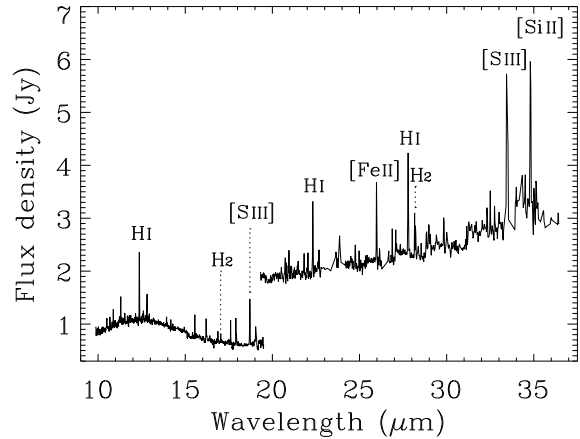}
\caption{IRS high-resolution spectra (modules SH and LH) of the
  central object. The most prominent spectral lines are labeled in the
  figure.
\label{Star-high}}
\end{figure}

\section{Results}

\subsection{Infrared images}

The distance assumed for \G\ is consistent with its probable
identification as a member of \objectname{Cyg OB2}. Located at the
southern part and close to the border of Cyg OB2, \G\ is behind the
Great Cygnus Rift, just to the northwest of the intense radio source
\objectname{DR 15}, which causes high visual extinction from 4 to at
least 10\,mag.  Thus, the infrared emission of \G\ is expected to be
heavily contaminated by foreground/background emission, as it is
clearly noticeable in all Spitzer images. In addition, a large-scale
infrared emission is also present toward the east of the images. This
large-scale emission increases its relative intensity with respect to
the background with the wavelength, and is similar to the large-scale
CO gas emission noted in Paper\,I at millimeter and submillimeter
wavelengths. It was related to the cloud around DR\,15
\citep{Oka01}. Due to this strong contamination, we refrain from
computing the dust mass, as it is a hard task and the results might
not be reliable enough.

The central object is detected in all Spitzer bands but MIPS
70\,\mic. We used MOPEX supported software to perform point-spread
function (PSF) photometry in the MIPS 24\,\mic\ image, obtaining a
flux density of 0.89\,Jy for the central object. Due to saturation
problems, it was not possible to obtain the fluxes of the central
object at shorter wavelengths.

The ring nebula previously detected by the {\it IRAS} and Midcourse
Space Experiment ({\it MSX}) satellites
\citep{Waters96, Egan02} is discernible from 3.6 to 70\,\mic. The
nebular emission dominates the image at 24\,\mic, being just above the
background level at shorter and longer wavelengths.  The nebula
presents a pretty spherical shape with a diameter of $\sim$\,3\farcm7
($\sim$\,1.8\,pc).  The shell looks like incomplete to the south,
probably due to a foreground IRDC, which is known to possess a very
large visual extinction \citep{Redman03}. However, it is not clear
whether this IRDC is part of the large complex including DR\,15 or it
is an unrelated foreground cloud \citep{Egan98}.

Thanks to the high angular resolution and sensitivity of these Spitzer
images, we have discovered an outstanding new, outer shell with almost
spherical shape as well. This outer shell is only visible in the
24\,\mic\ image (Figure~\ref{IRAC-MIPS}) and has a diameter of
$\sim$\,6\farcm8, ($\sim$\,3.4\,pc). During the process of writing
this paper, \cite{Gvaramadze09} reported a diffuse halo associated
with \G, which corresponds to the inside of this newly discovered
shell.

\subsection{CO maps}

As we show below, the CO data shown here improve the results gathered
in Paper\,I, due to a higher angular resolution and lower rms noise
compared to those in Paper\,I, providing a higher signal-to-noise
ratio (S/N). In addition, mapping the CO $J$\,=\,4\,$\rightarrow$\,3
line emission allows us to test a wider range of excitation
conditions, due to the higher energy levels involved and the critical
density needed to excite this line. According to the aim of this work,
here we provide the most relevant morphological features, postponing a
more detailed dynamical study to a future paper (J.~R.~Rizzo et
al.~2010, in preparation).

\begin{figure*}
\begin{center}
\includegraphics[angle=-90,width=0.85\textwidth,height=!]{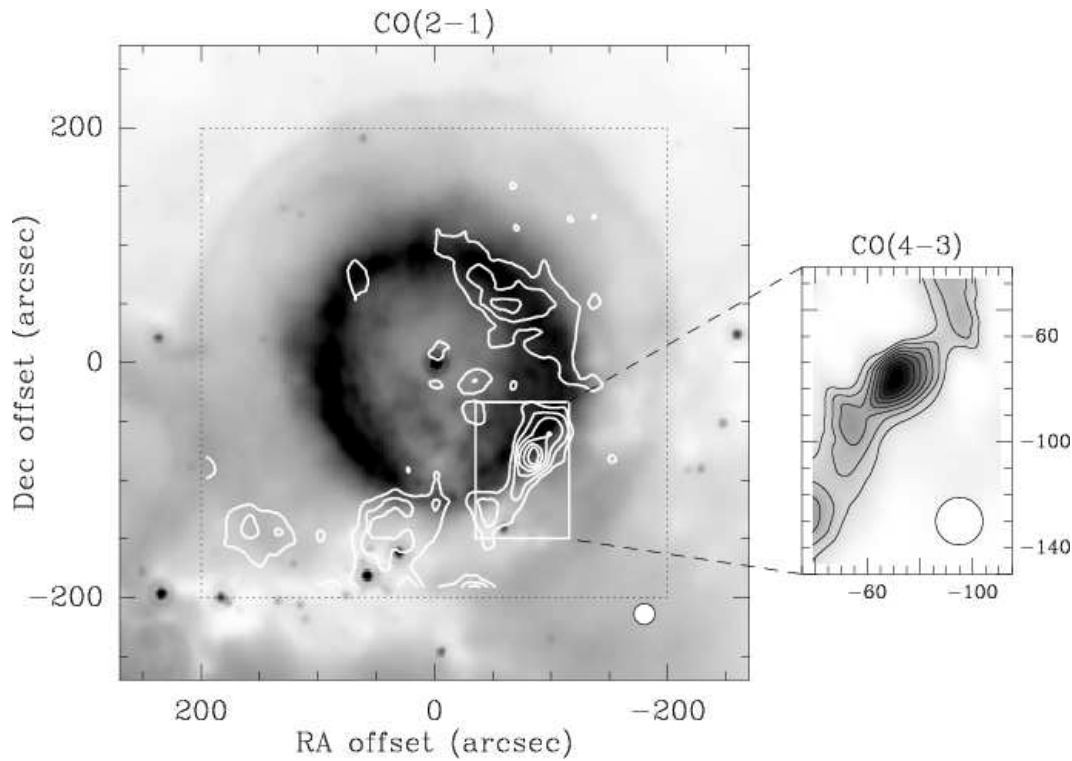}
\caption{{\em Left:} CO $J$\,=\,2\,$\rightarrow$\,1 map in the
  surroundings of \G, as obtained with the IRAM 30m radio
  telescope. The map is centered on the LBV star, and the velocity
  range of integration is from --3.2 to --2.6\,\kms. The pointed
  square indicates the observed field, and gray scale corresponds to
  the 24\,\mic\ emission. Contours are 40\,\%\,--\,90\,\% of the peak
  value (11.8\,K in $T_{\rm A}^*$ scale), in steps of 10\,\%. {\em
  Right:} CO $J$\,=\,4\,$\rightarrow$\,3 map in the same velocity
  range, corresponding exactly to the inset in the left panel, as
  obtained with the SMT 10\,m radio telescope. The inset has a size of
  80\arcsec$\times$116\arcsec. Contours are 25\,\%\,--\,95\,\% of the
  peak value (1.8\,K in $T_{\rm A}^*$ scale), in steps of 10\,\%. In
  both maps, the angular resolutions are shown as white circles in the
  lower right. It is striking the high correlation of the CO emission
  in this velocity range with the infrared inner shell. The CO shell
  is especially thin in the $J$\,=\,4\,$\rightarrow$\,3 map,
  unresolved by the beam.
\label{CO-}}
\end{center}
\end{figure*}

\begin{figure*}
\begin{center}
\includegraphics[angle=-90,width=0.85\textwidth,height=!]{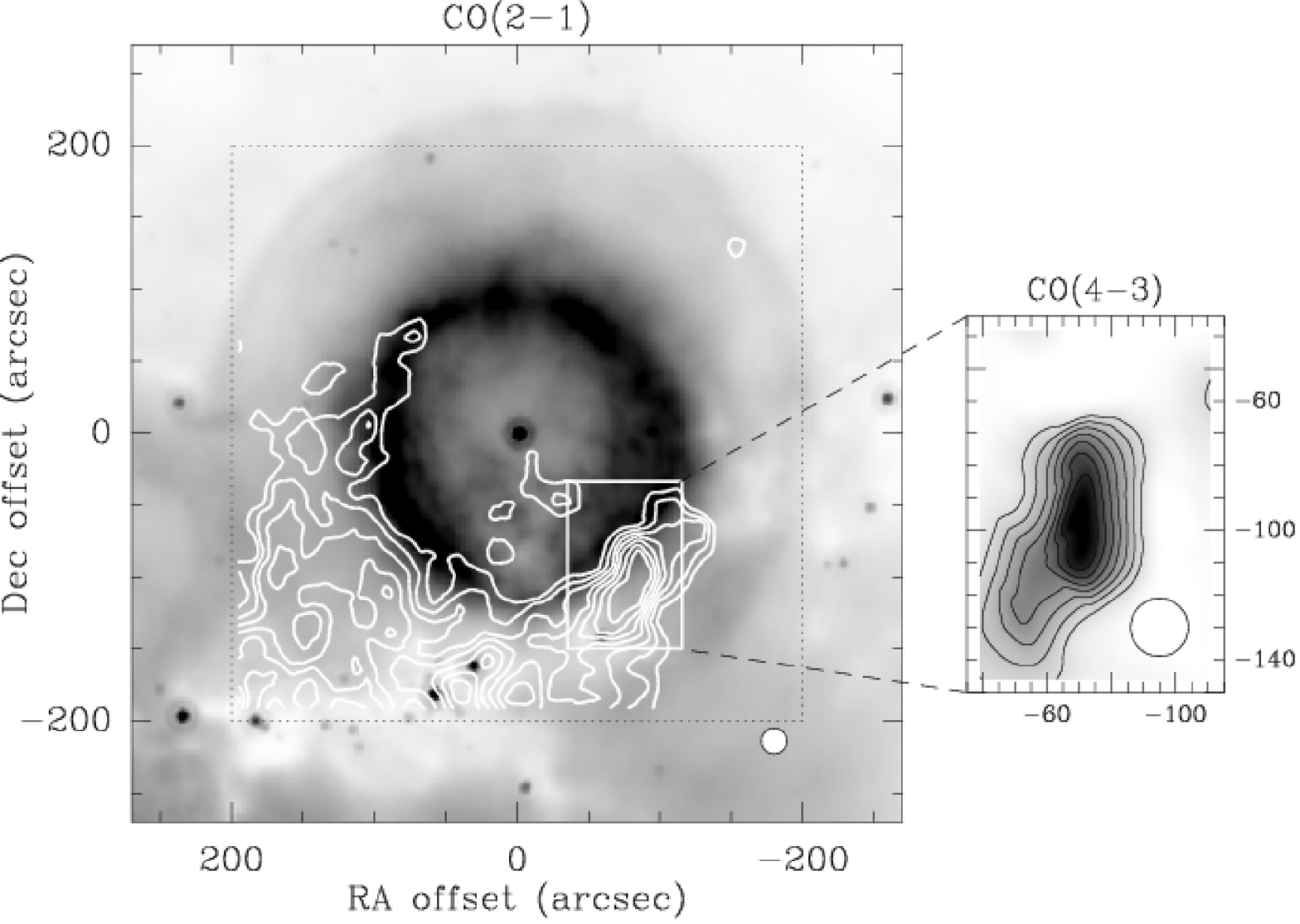}
\caption{Same as Figure~\ref{CO-}, but in the velocity range from
  +3.4 to +4.0\,\kms.  {\em Left:} contours are 30\,\%\,--\,90\,\% of
  the peak value (10.8\,K in $T_{\rm A}^*$ scale), in steps of
  10\,\%. {\it Right:} contours are 25\,\%\,--\,95\,\% of the peak
  value (3.4\,K in $T_{\rm A}^*$ scale), in steps of 10\,\%. This
  velocity range corresponds to a shocked clump reported in
  Paper\,I. After increasing the angular resolution, a second, outer
  shell is clearly visible in both lines.
\label{CO+}}
\end{center}
\end{figure*}

The whole velocity range of emission in the $J$\,=\,2\,$\rightarrow$\,1
line goes from --25 to 15\,\kms. The most intense emission corresponds
to the Local Arm, where the confusion effects are presumably the
highest. The most outstanding features in relation to \G\ appear in
velocity ranges similar to those shown in Paper\,I (see Figure~4 of
that paper). Figures~\ref{CO-} and \ref{CO+} show the CO
$J$\,=\,2\,$\rightarrow$\,1 and 4\,$\rightarrow$\,3 distributions in the
velocity ranges from --3.2 to --2.6 and from +3.4 to +4.0\,\kms,
respectively.  The CO $J$\,=\,2\,$\rightarrow$\,1 line contours are
superposed to an image at 24\,\mic\ in gray scale.

For the $J$\,=\,2\,$\rightarrow$\,1 data, a Gaussian convolution has
been applied, down to a resolution of 18\arcsec. This convolution
reduces the noise map and equals the angular resolution of the
$J$\,=\,4\,$\rightarrow$\,3 line. In both figures, the inset in the
$J$\,=\,2\,$\rightarrow$\,1 maps corresponds exactly to the mapped area
in the $J$\,=\,4\,$\rightarrow$\,3 line, shown on the right.

In Figure~\ref{CO-}, an arc-like feature dominates the map, in good
coincidence with the 24\,\mic\ inner shell, especially in the western
half.  The $J$\,=\,4\,$\rightarrow$\,3 line emission depicts part of a
remarkable thin shell, unresolved even at 18\arcsec\ of resolution.

The shocked clump, unresolved in Paper\,I, is clearly seen in
Figure~\ref{CO+}. In this velocity range, all the emission is outside
the 24\,\mic\ inner shell. The southwestern feature, also observed in
the $J$\,=\,4\,$\rightarrow$\,3 line emission, looks like part of a
second CO shell. Note that this feature is located outside the shell
shown in Figure~\ref{CO-}.

The emission at velocities intermediate to those of Figures~\ref{CO-}
and \ref{CO+} does not show any shell-like pattern, and consequently
we think that the two shells are different features, and not the
consequence of a mere velocity gradient.  Summarizing, at the
southwestern part of the field, we see two CO concentric slabs, which
fill the space between the two shells detected at 24\,\mic.

\subsection{Infrared spectra \label{IR-spec}}

The infrared spectral data show a rich variety of solid-state features
and narrow emission lines, superimposed on a thermal continuum.

Using the low-resolution IRS spectra, it is possible to study the
spectral energy distribution (SED) of the three observed targets
(Figure~\ref{all-low}). The SED of the central object (upper panel)
shows that the flux decreases from 5 to about 19\,\mic\ and increases
again to longer wavelengths. This behavior may be roughly explained by
two emitters at two different temperatures, i.e., hotter (shorter
wavelengths) and colder material (longer wavelengths). The middle and
lower panels of Figure~\ref{all-low} correspond to the spectra of the
shell and the IRDC, respectively. These two last SEDs are dominated by
the colder component.

The three SEDs present an astonishing similarity for
$\lambda$\,$>$\,25\,\mic. This suggests that this emission is due to
interstellar material which is present in all regions and extends
beyond the observed area.

Figure~\ref{AORs} shows the position of the IRS spectra slits for the
three regions observed. It is clear that the SL and LL spectra of the
central object are affected by the shell emission. The effect is
maximum at longer wavelengths, since the emission of the shell at
shorter wavelengths is negligible when compared with the emission of
the central object (see Figure~\ref{all-low}). Thus, in order to
remove the contribution of the shell emission to the central object
SED, we have scaled the LL spectrum of the shell to fit that of the
central object at the longest wavelengths ($>$\,32\,\mic) and later
subtracted from the last one. The result is shown in
Figure~\ref{all-low} (upper panel) with a solid line. The subtracted
central object SED shows the decreasing tail of the central object
emission. In addition, several spectral lines ($[$\ion{Ne}{3}$]$ at
15.5\,\mic, $[$\ion{S}{3}$]$ at 18.71 and 33.4\,\mic, H$_{2}$ at
28.2\,\mic, and $[$\ion{S}{3}$]$ at 34.8\,\mic) have disappeared of
the central object spectrum.

\subsubsection{Solid-state Features}

The SEDs of the central object and the IRDC reveal an absorption
feature at around 9.7\,\mic, more prominent in the case of the central
object, particularly clear after subtracting the shell contribution
(Figures~\ref{all-low} and \ref{short-low}). It corresponds to the
well-known 9.7\,\mic\ amorphous silicates feature due to the Si--O
stretching in the dust grains. The presence of amorphous silicates
was also reported in the dusty nebula surrounding the LBV stars
\objectname{HR Car} \citep{Lamers96a, Umana09} and \objectname{Wra
  751} \citep{Voors00a}. Note, however, that these stars present the
feature in emission instead of absorption as in the case of \G, which
is puzzling and has to be explained. In addition, the SED of the shell
seems to present a shallow depression at around 9.7\,\mic\ that could
be interpreted as a weak absorption due to amorphous silicate dust.

The broad, strong emission features at 6.2, 7.7, 8.6, 11.3, and
12.7\mic\ present in the SL spectrum of the shell (See
Figure~\ref{short-low}, upper panel) are due to vibrational modes of
free-flying PAHs (\citealt{Leger84, Allamandola85}). The same features
seem to appear in the spectrum of the IRDC, but the flux is too low
for reliable detections. These PAHs features suggest the presence of
very small C rich dust grains or molecules in the shell of \G. PAHs
have also been detected in the shell of the galactic LBV stars
AG\,Car \citep{Voors00a} and \objectname{HR 168625} \citep{Skinner97},
and in the LBV star \objectname{R71} of the Large Magellanic Cloud
\citep{Voors99}.

Another interesting solid-state feature, present only in the spectrum
of the IRDC, is the CO$_{2}$ bending mode at about 15\,\mic\ (see
Figure~\ref{all-low}, lower panel). Since the formation and evolution
of interstellar ices is strongly dependent on local conditions, the
presence of CO$_{2}$ ice in the spectrum of the IRDC indicates that
the physical conditions are rather different from those in the central
object and in the shell. CO$_{2}$ ice has never been detected in the
surroundings of a massive evolved star, but is usual in other
astronomical objects, such as high- and intermediate-mass protostars
\citep{Ehrenfreund98, Peeters02, Lefloch08}. The CO$_{2}$ absorption
feature found could be coming either from the quiescent regions of the
IRDC (south of the LBV nebula) or from the compact source lying within
the southern slit (see Figure~\ref{AORs}). Note that this compact source
was reported as a young stellar object by \cite{Redman03}.

\subsubsection{Emission lines}

The IRS high-resolution spectra of the central object are full of fine
structure lines that may provide important constraints to the physical
conditions in the region. The most prominent ones are also detected in
the low-resolution spectra and are labeled in the figures. For each
line, a first-order baseline has been defined and the line has been
fitted to a Gaussian. To do this, we have used an iterative software
developed by \cite{Bayo09}. The identification and the measured fluxes
are summarized in Table\,\ref{tab:lines}, including possible multiple
identifications. Flux errors are between brackets. All line fluxes for
the central object have been measured in the high-resolution IRS
spectra but \ion{H}{1} at 7.46\,\mic. In the case of the shell and the
IRDC, line fluxes have been measured in the respective low-resolution
IRS spectra. There is a mismatch between the SH and LH spectra of the
central object which is usually attributed to a difference in the two
module apertures, and which has been taken into account when comparing
fluxes of SH with those of LH spectra.

\begin{deluxetable}{ccccc}
\tabletypesize{\footnotesize}
\tablecaption{Line Identification and Measured Fluxes \label{tab:lines}}
\tablewidth{0pt}
\tablehead{
\colhead{Line} & \colhead{$\lambda^{a}$} &                              & \colhead{Flux (10$^{-14}$\,erg\,cm$^{-2}$\,s$^{-1}$)} &                  \\
               & \colhead{(\mic)}        & \colhead{The Central Object} & \colhead{The Shell}                                & \colhead{The IRDC}
}
\startdata
$[$\ion{Ar}{2}$]$; $[$\ion{Ne}{2}$]$                        &  6.97 &           & 7.7 (0.6)   & \\
\ion{H}{1} 6-5 8-6                                          &  7.46 & 960 (90)  &             & \\
$[$\ion{Ar}{3}$]$                                           &  8.99 &           & 6 (2)       & \\
\ion{He}{2} 10-9                                            &  9.67 &           &             & 2.6 (0.8) \\
\ion{He}{2} 24-16; \ion{H}{1} 12-8; $[$\ion{S}{4}$]$        & 10.50 & 19 (3)    &             & \\
$[$\ion{Ni}{2}$]$                                           & 10.68 & 7.7 (0.6) &             & \\
$[$\ion{Ne}{3}$]$                                           & 10.88 & 16 (2)    &             & \\
\ion{He}{2} 23-16 18-14; $[$\ion{Ni}{1}$]$; \ion{H}{1} 9-7  & 11.30 & 28 (2)    & 58 (8)$^{b}$ & \\
\ion{H}{1} 15-9                                             & 11.53 & 8 (2)     &             & \\
\ion{He}{2} 14-12; \ion{H}{1} 7-6                           & 12.37 & 77 (7)    &             & 1.9 (0.2) \\
$[$\ion{Ne}{2}$]$	                                    & 12.81 & 18 (2)    & 98 (8)$^{b}$ & 4.6 (0.2)$^{b}$ \\    
\ion{H}{1} 17-10; \ion{H}{1} 11-24                          & 13.93 & 6.7 (1.0) &            & \\ 
\ion{H}{1} 13-9                                             & 14.17 & 5.1 (0.6) &            & \\
\ion{H}{1} 23-11                                            & 14.29 & 3.0 (1.0) &            & \\
\ion{H}{1} 16-10                                            & 14.94 & 6.7 (1.3) &            & \\
$[$\ion{Ne}{3}$]$                                           & 15.55 & 13 (2)    & 29 (3)     & \\
\ion{He}{2} 20-16; \ion{H}{1} 10-8                          & 16.20 & 10 (2)    &            & \\
\ion{He}{2} 24-18; \ion{H}{1} 12-9                          & 16.87 & 3.6 (0.3) &            & \\
H$_{2}$ 0-0 S(1)                                            & 17.03 & 4.0 (0.5) & 19 (6)     & \\
\ion{H}{1} 18-11                                            & 17.60 & 6.7 (2)   &            & \\
$[$\ion{Fe}{2}$]$                                           & 17.90 & 8.4 (0.4) &            & \\
\ion{S}{3}                                                  & 18.71 & 24.0 (1.0)& 184 (15)   & 18 (4) \\
\ion{He}{2} 23-18 16-14; \ion{H}{1} 8-7                     & 19.05 & 24 (5)    &            & \\
\ion{H}{1} 20-12                                            & 20.52 & 4.5 (0.6) &            & \\
$[$\ion{Ar}{3}$]$                                           & 21.82 & 5.2 (0.6) &            & \\
\ion{He}{2} 17-15; \ion{H}{1} 13-10 11-9                    & 22.32 & 27 (2)    &            & \\
\ion{H}{1} 15-11                                            & 23.86 & 27.0 (1.4)&            & \\
$[$\ion{F}{1}$]$  	                                    & 24.75 & 9.0 (0.5) &            & \\
\ion{H}{1} 21-13                                            & 24.98 & 4.1 (0.8) &            & \\
$[$\ion{Fe}{2}$]$                                           & 25.97 & 23 (3)    &            & \\
\ion{H}{1} 24-14                                            & 27.08 & 7.3 (0.5) &            & \\
\ion{He}{2} 18-16; \ion{H}{1} 9-8                           & 27.79 & 29 (2)    &            & \\
H$_{2}$ 0-0 S(0)                                            & 28.21 & 13 (4)    &  25 (6)    & 11 (2) \\
\ion{H}{1} 14-11                                            & 28.85 & 7.7 (0.8) &            & \\
\ion{H}{1} 19-13                                            & 28.97 & 14 (3)    &            & \\
\ion{He}{2} 24-20; \ion{H}{1} 12-10                         & 29.83 & 10.3 (0.7)&            & \\
$[$\ion{S}{3}$]$	                                    & 33.45 & 78 (11)   & 740 (40)   & 229 (15) \\
$[$\ion{Si}{2}$]$	                                    & 34.80 & 41.5 (1.1)& 150 (40)   & 66 (6) \\
\enddata
\tablenotetext{a}{Peak wavelength measured on the stellar spectrum.}
\tablenotetext{b}{Flux measured affected by PAHs emission.}
\end{deluxetable}

The collection of spectral lines found is similar to that shown by the
LBV star HR\,Car \citep{Umana09} and the LBV candidate \objectname{HDE
  316285} \citep{Morris08}. The nebular spectra of HDE\,316285 show,
for the first time in this kind of objects, pure rotational lines of
H$_{2}$ arising from optically thin quadrupole transitions, from S(0)
to S(6). We have also clearly detected the two lower rotational lines,
S(0) and S(1), in the spectra of the central object and the shell, and
S(0) in the spectrum of the IRDC. Although these lines may arise from
a foreground component, or even DR\,15, the presence of H$_{2}$ in the
shell or in the circumstellar material is a possibility which has to
be explored in follow-up observations.

\section{Discussion}

\subsection{Multiple shells in \G}

The data shown in this paper clearly show multiple dust and gas shells
connected to the central star of \G. The MIPS 24\,\mic\ image shows
two well-defined dust shells. Figure~\ref{CO-} shows how the CO shell
coexists next to the inner dusty shell, while Figure~\ref{CO+} shows
another slab located between the two dust shells.

To explore the possibility of the existence of a more inner shell
closer to the LBV star, we have roughly fitted the SED of the central
object after subtracting the contribution of the shell with a
combination of two blackbodies. Two ranges of temperatures were tried,
from 10,000 to 30,000\,K for the stellar contribution and from 200 to
1500\,K for a possible inner envelope. Fitting does not vary
significantly in the considered ranges of temperature corresponding to
the hotter blackbody. However, the best fit regardless the stellar
temperature was always obtained with a second blackbody of
$\sim$\,900\,K. This result is compatible with the existence of an
inner shell unresolved by the Spitzer images. Figure~\ref{all-low}
shows the best fit for a stellar temperature of 20,000\,K. This
conclusion is supported by the detection of H$_2$ lines in the spectra
of the central object. It is necessary a thick dust envelope to be
able to explain the existence of H$_2$ so close to the LBV star. This
thick dust envelope would shield the molecule gas from the strong UV
field of the star.

From the morphology, it seems that the two dusty shells resolved by
the Spitzer images are interacting with the surrounding medium (see
Figure~\ref{IRAC-MIPS}), presumably with material belonging to the
DR\,15 cloud. Figure~\ref{CO+} also shows clear evidence of the
erosion of the ISM by the stellar wind, specially at the southwestern
region. At this region, we already reported in Paper\,I the presence
of a shock front.

In Paper\,I, we estimated an age of the nebula between $10^3$ and
$10^4$\,yr. This is in well agreement with the high sphericity shown
by the two dust shells. Thus, the mass-loss events which have produced
the multiple shells have occurred relatively recently, i.e., in the
current evolutionary stage.

In order to obtain a dust temperature, we have also fitted the
continuum emission of the shell and the IRDC, with a combination of
two blackbodies at different temperatures. In the case of the shell, a
blackbody of 65\,K and another of 132\,K were necessary. For the IRDC,
one of 66\,K and another of 187\,K were used. The resulting fits are
shown in Figure~\ref{all-low}.
In both cases, two dust components at different temperatures were
found, one common component at $\sim$\,65\,K and another
hotter. Probably, the cold component is due to foreground interstellar
material belonging to the \ion{H}{2} region DR\,15, as typical dust
temperatures of \ion{H}{2} regions are around 60\,--\,70\,K
\citep{Peeters02}. This is the main dust component and would explain
why both spectra have the same overall shape at longer
wavelengths. Regarding the hot component of the shell, its temperature
is similar (same order) to the temperatures estimated for the dust in
other LBV nebulae (300\,K; \citealp{Voors99}) and in supernova
remnants (SNRs) associated with shocks ($\sim$\,110\,K;
\citealp{Tappe06, Ghavamian09}). Note that the temperature of the hot
component of the shell is slightly lower than the temperature of the
hot component of the IRDC. Although these temperatures should be seen
with caution due to the presence of the CO$_2$ ice absorption feature,
the slightly higher temperature in the IRDC could be due to the
presence of a young stellar object in the slit (see Figure~\ref{AORs}).

The most external dust shell is only visible in the
24\,\mic\ image. It is expected that the dust in this shell would have
a lower temperature than the dust in the inner shells. Thus, the dust
temperature of the most external shell should not be well above
120\,K. Assuming that the mid-infrared images are tracing the thermal
continuum emission of the dust, this would explain why it is not
detected at shorter wavelengths. The fact that this shell is neither
detected at 70\,\mic\ could be due to a combination of strong
foreground/background emission at this wavelength, lower sensitivity
of MIPS in this band, and a non-appropriate data reduction of the
image.

The 9.7\,\mic\ amorphous silicates feature in absorption is clearly
present in the SL spectra of the central object and the IRDC, and is
tentatively detected in the shell. Amorphous silicates in LBV stars
are supposed to be formed at temperatures well below the silicates
melting point in the outer layers of the star which cool very rapidly
while the material is being swept up by a fast wind. The silicates
would indicate that the stellar ejecta was oxygen rich (C/O\,$<$\,1),
as expected for LBV stars which are supposed to have ejecta enriched
by the CNO cycle \citep{Nota97}. However, \G\ shows the
9.7\,\mic\ amorphous silicates feature in absorption instead of in
emission as the others LBV stars do. A plausible explanation is that
the absorption of amorphous silicates is due to the foreground ISM, as
was recently found in the case of the \objectname{SNR N157B} by
\cite{Micelotta09}. This would explain the presence of this feature in
absorption in the IRDC as well. However, another possible explanation
for the absorption shown in the central object spectrum is the presence
of amorphous silicate dust in an inner nebula unresolved by the
Spitzer images.

The presence of crystalline silicates in LBV nebulae seems to be a
general trend \citep{Waters97}. This is suggestive of the grains being
formed in high-density and low-temperature environments
\citep{Voors99}. Therefore, and because these features resemble those
observed in red supergiant stars, the presence of crystalline
silicates is explained if the dust is formed in high-density,
slow-expanding environments close to the star, rather than in the
swept-up shell. This kind of environment is typical of the red
supergiant phase \citep{Waters98}, for which wind velocities are
between 20 and 30\,\kms. In the case of \G, and similarly to the LBV
candidate HDE\,316235 \citep{Morris08} and the LBV stars
$\eta$\,Carina and HR\,Car \citep{Umana09}, its spectrum does not show
any signature of crystalline silicates. The lack of crystalline
silicates could be explained if the nebular material was rapidly
ejected and could not crystallize under conditions of rapid cooling
and low monomer densities, as suggested by \cite{Morris08} for
HDE\,316235. Another possible explanation would be amorphization of
initially crystalline dust, which may occur under heavy proton
bombardment due to the stellar wind. This last mechanism also explains
the lack of crystalline silicates in the ISM
\citep{Carrez02,Brucato04}.

\subsection{Features of ionized nebulae}

Ratios of fine structure lines can be compared with typical values
found toward \ion{H}{2} regions, photo-dissociation regions (PDRs),
SNR, and the ISM. First, we have considered the ratio between
$[$\ion{Si}{2}$]$\,34.8\,\mic\ and
$[$\ion{S}{3}$]$\,33.5\,\mic. Theseq two lines are found in the ISM,
and their ratio is indicative of the degree of excitation of the
gas. According to \cite{Simpson07}, this line ratio is $\sim$\,2.5 for
the diffuse interstellar gas, and just in the range 0.1\,--\,0.6
toward \ion{H}{2} regions. Since the values estimated for the \G\
region are similar to those of \ion{H}{2} regions (0.3 for the IRDC,
0.2 for the shell, and 0.5 for the central object), this suggests that
the three components studied in this work have important amount of
ionized gas. While this was expected for the shell and the central
object, where free--free thermal emission was already detected
\citep{Higgs94}, we note that the IRDC component is most likely
picking some emission from the \ion{H}{2} region DR\,15, which is
located only 4\arcmin\ to the (south) east of the LBV star and is
covered by the IRS slit.

Similarly, a correlation was found for the ratio between
$[$\ion{Ne}{3}$]$ 15.5\,\mic\ and $[$\ion{Ne}{2}$]$ 12.8\,\mic, and
the ratio between $[$\ion{S}{4}$]$ 10.5\,\mic\ and $[$\ion{S}{3}$]$
18.7\,\mic\ \citep{Martin-Hernandez02}. Such a correlation can be used
to measure the degree of ionization of a region. For the case of the
central object, the first ratio is around 0.7, and the second is
around 0.8 (assuming no contamination from \ion{He}{2} and \ion{H}{1}
at the $[$\ion{S}{4}$]$ line). By placing this value in the diagram
($[$\ion{Ne}{3}$]$/$[$\ion{Ne}{2}$]$~versus~$[$\ion{S}{4}$]$/$[$\ion{S}{3}$]$)
of Figure~1 of \cite{Martin-Hernandez02}, we have found that the
central object lies in an intermediate case between the \ion{H}{2}
regions with low and high ionization, and in particular it is close to
the value of the Orion Nebula. In addition, we have measured the ratio
between [\ion{Ne}{3}] 15.5\,\mic\ and $[$\ion{Ne}{2}$]$ 12.8\,\mic\
for the shell and have found a value around 0.3. It is interesting to
note the finding of \cite{Ghavamian09} that slow and dense shocks
produce a ratio around 0.4, which is very similar to the value found
in the shell of \G. This suggests the presence of slow shocks in the
shell, which is completely consistent with the recent discovery of a
slow CO shock in the southwestern side of the shell of \G\ (Paper\,I).

Finally, from the ratio between $[$\ion{S}{3}$]$\,33.4\,\mic\ and
$[$\ion{S}{3}$]$\,18.7\,\mic, one can make a rough estimate of the
electron density. The ratio\footnote{To compute this ratio, we have
  corrected the line intensity at 18.7\,\mic\ by a factor of 3. This
  factor for aperture losses applies only to the central object, for
  which the flux densities have been calculated in the high-resolution
  spectra, and the line at 18.7\,\mic\ falls in a different module
  than the line at 33.4\,\mic. For the shell and the IRDC, there is no
  need of correction because both lines fall in the same module.} is
1.1, 4.0, and 13.0 for the central object, the shell, and the IRDC,
respectively. The ratio of $\sim$\,1 corresponds to an electron
density of $\sim$\,500 cm$^{-3}$ \citep{Rubin89,Rubin94}. These
values are consistent with the electron density of
$\sim$\,1000\,cm$^{-3}$ derived by \cite{Umana09} for the central
object of HR\,Car. For the shell and the IRDC, this ratio is too high
to be able to derive a reliable electron density. This could be
indicative of strong extinction for the shell and the IRDC cases, as
found for some positions in the galactic center by \cite{Simpson07}.

\subsection{PDR signatures}

PAH emission features are present in the SL spectrum of the shell.
This dust material was also detected in the LBV stars AG\,Car,
HR\,168625, R71 \citep{Voors00b,Skinner97,Voors99}. PAH emission is
supposed to be formed where shocks act upon cold molecular gas. This
is presumably the situation of \G, where the ionized gas is
interacting with the molecular material traced by CO. Note that the
presence of shocks is supported by the morphology (see
Figures~\ref{IRAC-MIPS}, \ref{CO-}, and \ref{CO+}) and by the
detection of the CO $J$\,=\,4\,$\rightarrow$\,3 line, which has a
critical density of at least several $10^4$\,cm$^{-3}$.

On the other hand, the spectra of the central object show
$[$\ion{Fe}{2}$]$ lines, whereas the same lines are not detected in
the IRDC spectra. These two observational findings would indicate that
the Fe abundance close to the star is higher than in the surrounding
ISM. In the ISM, Fe is highly depleted from the gas phase because of
condensation onto dust grains, but shocks would reprocess these dust
grains and release Fe atoms into the gas phase. Therefore, a high
abundance of Fe in comparison with the ISM would be another evidence
of the presence of shocks close to the LBV star.

The presence of PAHs in the spectrum of the shell, the detection of
low-excitation fine structure lines such as
26.0\,\mic\ $[$\ion{Fe}{2}$]$ in the central object spectrum, and the
CO gas emission, point out to the existence of a PDR surrounding the
ionized part of the nebula. \cite{Laurent00} showed a typical spectrum
of a star-forming region, an \ion{H}{2} region, and a PDR (see their
Figures~1 and 2). In our case, the spectrum of the shell shown in
Figure~\ref{short-low} is an intermediate case between a PDR and an
\ion{H}{2} region. In addition, the non-detection of $[$\ion{S}{1}$]$
at 25\,\mic\ and the detection of $[$\ion{S}{3}$]$ are characteristic
of photon-ionized gas \citep{Micelotta09}. In fact, $[$\ion{S}{3}$]$
and $[$\ion{Si}{2}$]$ are typical PDR tracers \citep{Tappe06,Umana09},
along with $[$\ion{Ar}{3}$]$, $[$\ion{Ne}{2}$]$, $[$\ion{Ne}{3}$]$,
and $[$\ion{S}{4}$]$
\citep[e.g.,][]{Martin-Hernandez02,Compiegne08,Ghavamian09}, all
present in the central object spectra.

In PDRs, PAHs are non-thermally excited by single ultraviolet (UV)
photons \citep{Sellgren84}, and far-UV photons control both the
thermal and chemical structure of the neutral gas \citep{Tielens85}.
The high values of both the stellar effective temperature ($T_{\rm eff}$) 
and luminosity ($L_*$) favor the presence of a PDR where the significant 
UV field encounters cold molecular gas. However, as the CO shells 
are at large stellar distances, the presumably high UV field may be 
diluted. 

Thus, we have made an attempt to estimate the incident UV field
($G_0$) in the 24\,\mic\ inner shell. $G_0$ is defined as the fraction
of an incident UV field in the ionizing UV range (i.e., between 6 and
13.6\,eV). This is usually expressed in units of the Habing field
($1.2\times10^{-4}$\,erg\,cm$^{-2}$\,s$^{-1}$\,sr$^{-1}$), which is a
representative value of the mean interstellar UV field. $G_0$ depends
on the stellar parameters as

\begin{equation}
G_0 \propto f_{\rm UV}(T_{\rm eff}) \cdot L_* \cdot d^{-2},
\end{equation}

\noindent
where $d$ is the distance of the affected volume to the star and
$f_{\rm UV}$ is the fraction of the Planck function emitted between 6
and 13.6\,eV with respect to the total:

\begin{equation}
f_{\rm UV}(T_{\rm eff}) = \frac{\int_{6\,eV}^{13.6\,eV}\!\!B_{\nu}(T_{\rm eff})}{\int_{0}^{\infty}\!\!B_{\nu}(T_{\rm eff})}.
\end{equation}

By varying $T_{\rm eff}$ between 15,000 and 30,000\,K, and $L_*$
between 1 and $2\times10^{5}$\,L$_\odot$, $G_0$ ranges from 0.8 to
$3.0\times10^4$, in units of Habing field.  In the assumed range of
$T_{\rm eff}$, $f_{\rm UV}$ varies from 31\% to 56\% of the total
energy emitted by the star. A value for $d$ of 1\,pc was assumed,
which corresponds to the distance of the 24\,\mic\ shell to the star,
at a distance of 1.7\,kpc.

These high values of $G_0$ (of the order of $10^4$), together with
moderately high H$_2$ densities (at least several $10^4$\,cm$^{-3}$
due to the presence of CO $J$\,=\,4\,$\rightarrow$\,3), are similar to
the conditions found in extreme PDRs, such as the well-known
\objectname{Mon R2} \citep{Rizzo03a,Berne09}. It is also compatible
with the values found by \citet{Umana09} in HR\,Car. However, PAHs are
expected to be not widespread but located in small regions where the
restrictive conditions of PDRs apply. A zone close to the inner
24\,\mic\ shell and the CO $J$\,=\,4\,$\rightarrow$\,3 shells is a good
candidate to search for these fragile large molecules.

\subsection{Molecular hydrogen}

The 0--0 S(0) line of H$_{2}$ has been detected in the three targets,
while the S(1) line has only been identified in the central object
and the shell. Unfortunately, we have not detected more excited pure
rotational lines (S(2) and above), which prevent us a thorough
excitation study. Despite that, the mere presence of molecular
hydrogen in the close environs of a LBV star is interesting and
deserves further analysis. To the best of our knowledge, the only
reported detection of H$_{2}$ close to a LBV was done by
\citet{Morris08}. It is important to note that \citet{Umana09}, using
similar data sets than ours, did not found any H$_{2}$ lines in the
dusty nebula HR\,Car.

Vibrationally excited H$_{2}$ was also discovered surrounding part of
the Wolf-Rayet nebula \objectname{NGC\,2359} \citep{St-Louis98}, in
correlation with significant amounts of (probably) shocked CO
\citep{Rizzo01,Rizzo03b}.

Our values of the S(1)/S(0) line ratio are not too different to those
found by \citet{Tappe06} in a molecular cloud to an SNR, affected by
both UV radiation and shocks. Unfortunately, a deeper analysis from
this line ratio is not appropriate due to the fact that both lines
correspond to different species, and we cannot constrain the
ortho-to-para ratio. A better study including more lines and a precise
spatial distribution may help to understand the nature of the emitting
H$_{2}$ lines, as well as a comparison to several PDR models
\citep[see, e.g.][]{Draine96,Compiegne08}

\section{Conclusions}

In order to shed light on the distribution of warm dust, PAHs and CO
gas in \G, we have analyzed infrared Spitzer data and observed the CO
$J$\,=\,2\,$\rightarrow$\,1 and 4\,$\rightarrow$\,3 lines using the IRAM
30m radio telescope and the SMT.

An analysis of the high-sensitivity MIPS 24\,\mic\ image has led us to
the discovery of a second outer dusty shell with a spherical shape,
previously overlooked. In addition, the analysis of the SED of the
central object has revealed the possible existence of a third
extremely young dust shell closer to the LBV star, unresolved in the
Spitzer images. Furthermore, the new CO data shown in this paper have
revealed two separated gaseous shells located between the two dusty
shells. The dynamic time of these gaseous components would indicate
that the various shells have been formed during the current
evolutionary stage. The presence of multiple shells around a LBV star
is of great interest because its implications for stellar evolution
theories. This would indicate that the mass-loss history has been more
complex than previously assumed, with periods of enhanced and dimmed
stellar winds instead of a continuous steady one.

Using the IRS low-resolution spectra, we have compared three regions in
the field of \G: the central object, the shell, and the IRDC. The three
studied regions present a clear similarity at the longest wavelengths
($>$\,25\,\mic), due to the presence of a large amount of
foreground/background dust at $\sim$\,65\,K. A rich variety of
solid-state features has been detected. Amorphous silicate features in
absorption are present in the spectra of the central object and the
IRDC, probably due to the foreground ISM and/or to the presence of
amorphous silicate dust in a likely inner nebula very close to the
central LBV star. PAHs features in emission are present in the
spectrum of the shell, and a tentative detection of amorphous silicate
in absorption is reported as well. Furthermore, a CO$_{2}$ ice feature
in absorption is present in the spectrum of the IRDC.

The IRS high-resolution spectra of the central object are full of fine
structure lines, similar to those detected in other LBV stars. We have
also detected pure rotational lines of H$_{2}$ arising from optically
thin quadrupole transitions in two regions of \G, the central object
and the shell, which is the first detection of H$_2$ in the spectrum
of a LBV star.

Using different flux line ratios, we have analyzed the physical
conditions of \G\ and its surroundings. In the case of the shell,
indications of slow shocks have been found.

We have estimated an incident UV field in the inner dusty shell of the
order of 10$^{4}$ in units of Habing field. This high value of $G_0$,
together with other evidence such as moderately high H$_2$ densities,
the presence of PAHs in the shell, the detection of low-excitation fine
structure lines, and the CO mid-J emission, suggests the existence of a
PDR.

This study of the multiple phase material directly associated with
\G\ may help to understand its recent past and consequently the
evolutionary mechanisms of this kind of object. In this sense,
\G\ becomes a paradigm of LBV nebula and deserves further studies. In
particular, a follow-up study of complex molecules should also be
carried out, because they can provide information about the physical
parameters in detail. In addition, similar studies to the one shown in
this paper for other LBVs and evolved massive stars will also help to
understand this short, but crucial, evolutionary phase of high-mass
stars.

\acknowledgments

The authors thank the referee for her/his comments and suggestions. We
are grateful to A. Bayo for her self-written software to analyze
spectral line fluxes and her friendly support, and to
\'A. S\'{a}nchez-Monge for useful discussions and modeling of the
centimeter emission. This work is based in part on observations made
with the {\it Spitzer Space Telescope}, which is operated by the Jet
Propulsion Laboratory, California Institute of Technology under a
contract with NASA. CO data reduction and analysis were done using the
GILDAS package (\url{http://www.iram.fr/IRAMFR/GILDAS}). The SMT is
operated by the Arizona Radio Observatory (ARO), Steward Observatory,
University of Arizona. This work is partially funded by the Spanish
MICINN under the Consolider-Ingenio 2010 Program grant CSD2006-00070:
First Science with the GTC
(\url{http://www.iac.es/consolider-ingenio-gtc}).


{\it Facilities:} \facility{Spitzer Space Telescope}, \facility{IRAM 30m}, \facility{SMT}.

\bibliographystyle{apj} 
\bibliography{/pcdisk/muller/fran/RESEARCH/bibliography/references.bib}

\end{document}